# Effects of Band Symmetry on Spin-Dependent Transport in Noncollinear Antiferromagnetic Tunnel Junctions

Mohamed Elekhtiar[1], Ding-Fu Shao[2], and Evgeny Y. Tsymbal[1*]

[1] *Department of Physics and Astronomy & Nebraska Center for Materials and Nanoscience, University of Nebraska, Lincoln, Nebraska 68588-0299, USA*

[2] *Key Laboratory of Materials Physics, Institute of Solid-State Physics, HFIPS, Chinese Academy of Sciences, Hefei 230031, China*

Antiferromagnetic tunnel junctions (AFMTJs) can exhibit large tunneling magnetoresistance (TMR), making them promising candidates for ultrafast and field-robust spintronic devices. Here, we elucidate the role of band symmetry in governing spin-dependent transport in AFMTJs. Using first-principles density-functional theory combined with quantum-transport calculations, we investigate $Mn_3NiN/LaAlO_3/Mn_3NiN$ (001) junctions based on the noncollinear $\Gamma_{4g}$ antiferromagnetic phase of $Mn_3NiN$. Although $Mn_3NiN$ exhibits a large momentum-dependent spin polarization due to broken $PT$ symmetry, we show that the tunneling conductance is critically controlled by band symmetry of the electrode Bloch states and their symmetry-selective coupling to evanescent states in the $LaAlO_3$ barrier. Orbital-symmetry selection rules suppress interband transmission in the parallel configuration, whereas the antiparallel configuration enables symmetry-compatible interband tunneling along the diagonal directions of the two-dimensional Brillouin zone. These additional transmission channels enhance the antiparallel conductance and reduce the TMR relative to predictions based solely on spin polarization. Nevertheless, the TMR remains exceptionally large, exceeding 2000%, while band symmetry controls the attainable magnitude of TMR in AFMTJs. Our results establish band-symmetry filtering as an essential ingredient of spin-dependent tunneling in AFMTJs.

*e-mail*: tsymbal@unl.edu

## I. INTRODUCTION

Spintronics exploits the electron's spin degree of freedom to realize functionalities beyond those achievable with conventional charge-based electronics [ 1 ]. To date, ferromagnetic materials have served as the primary active elements in spintronic devices, such as magnetic tunnel junctions (MTJs) [2]. This is due to their spontaneous magnetization and long-range magnetic order which enable efficient generation and detection of spin-polarized currents. These capabilities underpin well-established phenomena such as spin-dependent transport [3-5], tunneling magnetoresistance [6-8], and current-induced switching [9,10]. However, their finite stray fields, relatively slow spin dynamics, and magnetic crosstalk impose fundamental limitations on device scalability, speed, and energy efficiency.

Antiferromagnets offer a compelling alternative. Owing to their zero net magnetization, antiferromagnetic (AFM) materials generate no stray magnetic fields, eliminating magnetic crosstalk and enabling densely packed device architectures. In addition, their intrinsically ultrafast spin dynamics allow operation in the terahertz regime while remaining robust against external magnetic perturbations. Together, these characteristics make antiferromagnets particularly attractive for high-density, low-power spintronic applications, where stability, speed, and scalability are essential [11-15].

Despite their intrinsic advantages, controlling and detecting the AFM order parameter is fundamentally more challenging than in ferromagnets, where the net magnetization provides a direct and easily accessible handle for both electrical manipulation and readout. In conventional collinear antiferromagnets, the two magnetic sublattices are typically related by combined symmetries, such as $PT$ or $U\tau$, where $P$ is spatial inversion, $T$ is time reversal, $U$ is spin flip, and $\tau$ is lattice translation. These symmetries enforce complete compensation of the spin polarization and suppress magnetotransport responses. These challenges can be alleviated in noncollinear antiferromagnets, where the magnetic symmetry is reduced such that neither $PT$ nor $U\tau$ symmetry is preserved. As a result, such non-collinear antiferromagnets can exhibit ferromagnet-like functionalities, including electrically detectable transport signals and current-induced manipulation, while retaining intrinsic advantages of antiferromagnets, such as ultrafast dynamics and zero stray magnetic fields [16,17].

Specifically, non-collinear antiferromagnets enable a range of phenomena that are forbidden in conventional antiferromagnets, including the anomalous Hall effect [18-25], spin-polarized charge currents [26,27], and the AFM-order-dependent ($T$-odd) spin Hall effect [28-31]. These properties originate from the unique non-collinear magnetic order, which can provide efficient electrical and electromechanical pathways for probing and controlling the AFM order parameter. In addition, the magnetic state of non-collinear antiferromagnets can be manipulated electrically via spin-orbit torques [32-35] and can also be controlled through strain when integrated with piezoelectric substrates [36-38].

Especially exciting, however, is the recent theoretical prediction [39-41] and experimental demonstration [42-46] that AFM order can be harnessed in antiferromagnetic tunnel junctions (AFMTJs) producing a sizable tunneling magnetoresistance (TMR) effect. In contrast to conventional MTJs based on ferromagnetic electrodes, where TMR arises from the relative alignment of their net magnetizations, AFMTJs exploit the sensitivity of spin-dependent tunneling to the symmetry, orientation, and microscopic configuration of the AFM order parameter [47]. In particular, non-collinear antiferromagnets with broken *PT* symmetry, such as $Mn_3X$ and $Mn_3XN$ ($X$ = Ir, Sn, Ge, Pt), support a large momentum-dependent spin polarization [40,41] resulting in spin-polarized tunneling currents that depend on the AFM order parameter despite the absence of net magnetization. As a result, the resistance of an AFMTJ can be switched between distinct states associated with different AFM configurations, enabling nonvolatile electrical readout. These developments establish AFMTJs as a promising platform for AFM spintronics, combining robust TMR signals with ultrafast dynamics, high-density integration, and compatibility with energy-efficient electrical control.

To date, the physics of TMR in AFMTJs has been largely attributed to the momentum-dependent spin polarization of the AFM electrodes, which is a key factor underlying the effect [40,41,47,48-50]. However, it is well known that in crystalline MTJs, where the transverse wave vector is conserved during tunneling, an accurate description of spin-dependent transport must explicitly account for the symmetry matching between the Bloch states in the electrodes and the evanescent states in the tunneling barrier [51]. In particular, the efficient coupling of the majority-spin $\Delta_1$-symmetry band in bcc Fe (001) to the slowly decaying $\Delta_1$-symmetry evanescent state in the MgO (001) is responsible for a large positive spin polarization and giant TMR observed in Fe/MgO/Fe (001) MTJs [52]. Conversely, symmetry considerations also explain the large negative spin polarization of electrons tunneling from bcc Co(001) through an $SrTiO_3$ (001) barrier [53] consistent with the experimental observations [54,55]. These examples demonstrate that the transport spin polarization in MTJs is governed not by the magnetic electrode alone, but by the combined symmetry properties of the electrode-barrier pair, a concept that can be formally captured by the interface transmission function [56,57].

By the same reasoning, symmetry-selective tunneling is expected to play a central role in determining the transport properties of crystalline AFMTJs. Conservation of the transverse wave vector $\boldsymbol{k}_\parallel$ implies that the tunneling conductance depends not only on the $\boldsymbol{k}_\parallel$-dependent spin polarization of the AFM electrodes, but also on the symmetry matching between AFM Bloch states and evanescent states in the barrier. As a result, reorientation or reconfiguration of the AFM order parameter can modify the symmetry of the tunneling channels, leading to pronounced changes in spin-dependent transmission. This suggests that both the magnitude and sign of TMR in AFMTJs are controlled by the specific AFM-barrier combination, thereby extending the concept of symmetry-filtered tunneling beyond conventional ferromagnetic MTJs.

Motivated by these considerations, in this work we explore the role of band symmetry on the electronic structure and quantum-transport properties of crystalline AFMTJs. Based on first-principles density-functional calculations, we consider AFMTJs composed of the cubic antiperovskite $Mn_3NiN$ in its non-collinear $\Gamma_{4g}$ phase as the electrode material, together with closely lattice-matched $LaAlO_3$ as the tunneling barrier. This system provides a representative platform for studying symmetry-controlled tunneling in non-collinear AFMTJs. We find that spin-dependent electron transmission is governed by the combined effect of band symmetry intrinsic to the electrodes and symmetry-selective coupling between electrode Bloch states and evanescent states in the barrier. In particular, in $Mn_3NiN/LaAlO_3/Mn_3NiN$ (001) junctions, symmetry allows additional conduction channels in the low-resistance antiparallel configuration, enhancing its conductance and thereby reducing the resulting TMR. Nevertheless, the predicted TMR remains exceptionally large, exceeding 2000%, demonstrating that band symmetry plays a central role in determining the attainable magnitude of TMR in AFMTJs.

## II. METHODS

Calculations are performed within density functional theory (DFT) using the plane-wave pseudopotential method implemented in Quantum-ESPRESSO [58]. Ultrasoft pseudopotentials [59] and generalized gradient approximation (GGA) for exchange-correlation potential [60] are employed. A plane-wave kinetic cut-off energy of 64 Ry and a Gaussian broadening of 0.01 eV are used in all calculations. To simplify the calculations, the La 4*f*-electrons are treated within the frozen-core (*f*-in-core) approximation, as their contribution in the $La^{+3}$ oxidation state of $LaAlO_3$ is negligible within the energy window of interest. Electronic structures of bulk $Mn_3NiN$ and $LaAlO_3$ are converged using a 25×25×25 *k*-point grid and shown in Fig. A1 and Fig. A2, respectively. The calculated band structures are in good agreement with the previous theoretical studies [61,62].

The relaxed lattice parameters of bulk $Mn_3NiN$ and $LaAlO_3$ were found to be 3.84 Å and 3.81 Å, respectively, in close agreement with the corresponding experimental values of 3.87 Å [63] and 3.81–3.83 Å [64]. This corresponds to only ~2% lattice mismatch, indicating the feasibility of epitaxial growth of a heterostructure composed of these two compounds. In constructing the $Mn_3NiN/LaAlO_3/Mn_3NiN$ (001) junction, the in-plane lattice parameter was fixed to that of $Mn_3NiN$ to simulate epitaxial growth ($a$ = 3.84 Å). Under this constraint, relaxation of the out-of-plane lattice constant yielded $c$ = 3.804 Å for bulk $LaAlO_3$.

To determine the energetically most favorable structure of the $LaAlO_3/Mn_3NiN$ (001) interface, several interface terminations are considered and their formation energies and internal pressures are evaluated, as described in Appendix III. Structural relaxations of the heterostructures are performed using a 7×7×1 *k*-point mesh. The internal atomic coordinates and out-of-plane lattice constant are optimized until the residual forces on all atoms are below $2.5\times10^{-4}$ eV/Å and the residual stress is less than 0.5 kbar.

Table A1 summarizes the calculated interface formation energies and internal pressures for the different interface configurations. Although the LaO-MnNi interface termination exhibits the lowest formation energy, it retains an unphysically large residual internal pressure even after relaxation. Therefore, the $AlO_2$-$Mn_2N$ terminated interface, which has the second-lowest formation energy, was identified as the most stable physically meaningful configuration and was adopted for the main analysis presented in this paper. For this interface, the Mn-O bond length was found to be 2.124 Å, which is somewhat shorter than the bond length in bulk MnO in the rocksalt structure (~2.22 Å) due to altered coordination and bonding environment. Subsequent self-consistent electronic-structure calculations for the AFMTJ heterostructure were performed using a denser 10×10×1 $k$-point point mesh.

The spin expectation values of individual Bloch states and the spin polarization of the conducting channels are evaluated using a 50×50×150 $k$-point mesh. The layer-resolved spectral density (i.e. the layer- and $\boldsymbol{k}_\parallel$-resolved density of states) of the relaxed $Mn_3NiN/LaAlO_3/Mn_3NiN$ (001) junction is calculated with a supercell approach involving periodic boundary conditions and a 50×50 in-plane $k$-points mesh. The effects of spin-orbit coupling (SOC) on spin polarization of bulk $Mn_3NiN$ (001) and $Mn_3NiN/LaAlO_3/Mn_3NiN$ (001) AFMTJs are neglected.

Quantum-transport calculations are performed using the PWCOND code [ 65 , 66 ] implemented within Quantum ESPRESSO. The $Mn_3NiN/LaAlO_3/Mn_3NiN$ (001) junction is constructed from a fully relaxed $Mn_3NiN/LaAlO_3$ (001) heterostructure consisting of a periodic stack of 6.5 unit cells (u.c.) of $Mn_3NiN$ and 3.5 u.c. of $LaAlO_3$. This junction is treated as the scattering region ideally attached on both sides to semi-infinite $Mn_3NiN$ leads. Spectral and channel resolved transmissions are obtained using a 100×100 $k$-point mesh in the two-dimensional Brillouin zone (2DBZ). Total transmission as a function of energy is computed using the same $k$-point sampling.

For reliable transport calculations, it is essential that the layers at the interfaces between the central region and the semi-infinite leads be bulk-like; otherwise, artificial scattering may arise from unphysical electronic mismatch due to an insufficient electrode thickness. Additional care is required when constructing antiparallel magnetic configuration. In this case, both translational periodicity and magnetic compatibility must be preserved at the boundaries where the semi-infinite leads are attached, in order to avoid abrupt changes in the magnetic structure that would otherwise introduce spurious scattering and convergence issues. To address this, the heterostructure is doubled along the transport direction, creating two distinct electrode regions, each containing bulk-like $Mn_3NiN$ layers with the appropriate magnetic order. These regions are then independently connected to the corresponding semi-infinite leads, eliminating artificial boundary effects and ensuring a consistent and physically meaningful AFMTJ model.

The decay rates of evanescent states in $LaAlO_3$ are obtained from its complex band structure calculated using PWCOND. An arbitrary wave vector is decomposed into in-plane component $\boldsymbol{k}_\parallel$, which is conserved during tunneling, and an out-of-plane component $k_z$. For each $\boldsymbol{k}_\parallel$, the dispersion relation $E = E(k_z)$ is computed, allowing for the complex wave vectors $k_z = q + i\kappa$ . The imaginary part $\kappa$ defines the decay rate, such that the corresponding evanescent wave functions decay as $\sim e^{-\kappa z}$. The irreducible representations of the evanescent states are identified by continuously matching them to the real bands at $\kappa = 0$. The symmetry character of the real bands is determined from their orbital decomposition and the corresponding real-space symmetry representations [Figs. A1 and A2].

Figures are plotted using Matplotlib [67], Mathematica [68] and VESTA [69].

## III. RESULTS

### A. Atomic and magnetic structure of $Mn_3NiN$

$Mn_3NiN$ crystallizes in the cubic antiperovskite structure with space group $Pm\bar{3}m$, where Ni atoms occupy the cube corners, N resides at the body center, and Mn atoms form a face-centered sublattice at the $3c$ Wyckoff positions, as illustrated in Fig. 1(a). The Mn atoms surround the central N atom, forming an $Mn_6$ octahedron, while the Mn sublattice itself can be viewed as a stacking of kagome-like planes normal to the [111] direction. This geometrically frustrated Mn network provides the microscopic basis for stabilizing noncollinear AFM order.

In the noncollinear $\Gamma_{4g}$ magnetic phase, the three Mn moments within the primitive cell adopt a coplanar 120° configuration, as shown in Fig. 1(b). The calculated local magnetic moment is about 2.45 $\mu_B$/Mn atom, in qualitative agreement with available experimental measurements and previous first-principles calculations [63, 70 ]. The magnetic moments lie within the kagome planes and form a chiral triangular arrangement. This magnetic structure preserves the translational symmetry of the crystal but breaks $T$ symmetry and several spatial symmetries of the paramagnetic phase, while remaining compatible with a subset of rotational operations combined with time reversal, such as the $M_{(1\bar{1}0)}T$ symmetry indicated in Fig. 1(b). The resulting magnetic structure belongs to the magnetic space group $R\bar{3}m'$.

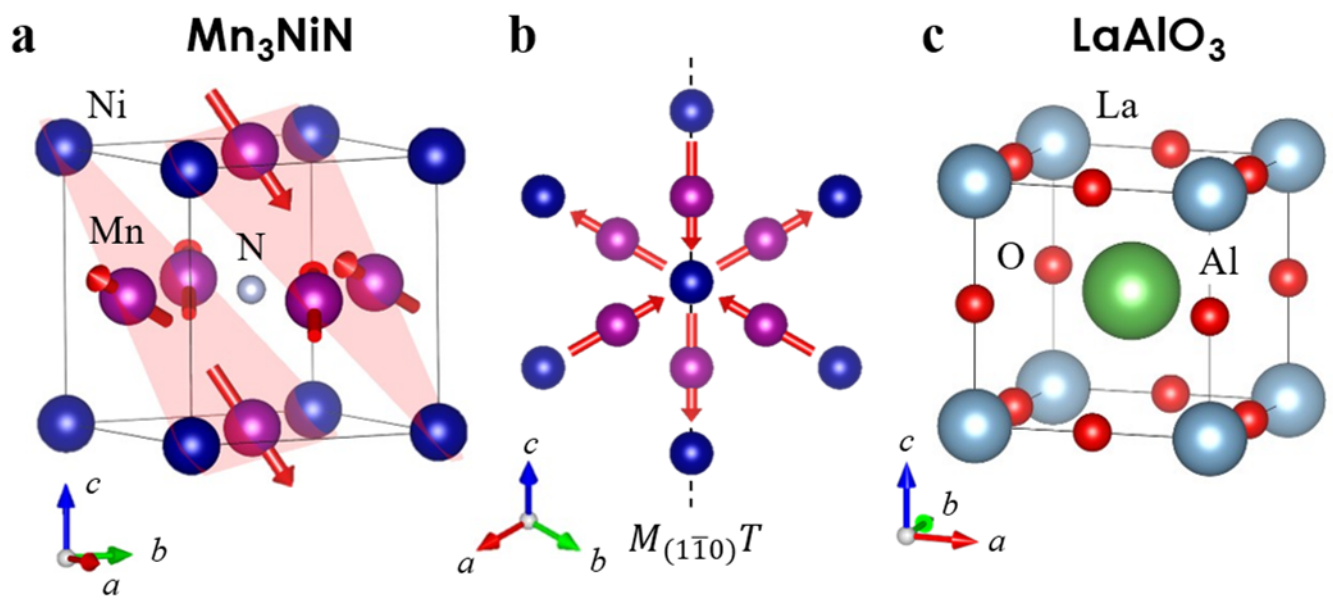

**Fig. 1. Structure of $Mn_3NiN$ and $LaAlO_3$.** (a,b) Atomic and magnetic structure of antiperovskite $Mn_3NiN$ in the noncollinear AFM $\Gamma_{4g}$ phase in a 3D view (a) and projected onto the (111) plane (b). The $M_{(1\bar{1}0)}T$ symmetry of the magnetic space group of $Mn_3NiN$ is indicated. (c) Atomic structure of perovskite $LaAlO_3$.

Although the underlying crystal lattice is centrosymmetric, inversion symmetry $P$ does not restore the magnetic structure when combined with time reversal. Each Mn site in the antiperovskite lattice is inversion invariant (up to a lattice translation), so the combined $PT$ operation reverses the direction of the local magnetic moment without mapping it onto a distinct Mn sublattice. As a result, the $\Gamma_{4g}$ spin configuration is not invariant under $PT$. This absence of $PT$ symmetry, a key characteristic of the AFM $\Gamma_{4g}$ phase, underlies the emergence of momentum-dependent spin polarization.

### B. Momentum-dependent spin polarization

A momentum-dependent spin polarization plays a central role in controlling the spin-dependent transport properties of noncollinear antiferromagnets with broken $PT$ symmetry, as well AFMTJs based on these materials. For a given transverse wave vector $\mathbf{k}_{\parallel}$, the spin state at the Fermi energy is defined in terms of the expectation value of the spin operator $\hat{\mathbf{s}}$ as

$$\boldsymbol{s}_{\boldsymbol{k}_{\parallel}} = \sum_{n} \boldsymbol{s}_{n\boldsymbol{k}_{\parallel}} = \sum_{n} \frac{l_z}{2\pi} \int \langle \psi_{n\boldsymbol{k}} | \hat{\mathbf{s}} | \psi_{n\boldsymbol{k}} \rangle \, \delta(E_{n\boldsymbol{k}} - E_F) dk_z \,. \quad (1)$$

Here $l_z$ is the lattice constant of the electrode along the transport $z$ direction, $\langle \psi_{n\boldsymbol{k}} | \hat{\mathbf{s}} | \psi_{n\boldsymbol{k}} \rangle$ is the spin expectation value for band $n$ with energy $E_{n\boldsymbol{k}}$ and eigenstate $\psi_{n\boldsymbol{k}}$ at wave vector $\boldsymbol{k} = (\boldsymbol{k}_{\parallel}, k_z)$, and $E_F$ denotes the Fermi energy.

Based on $\boldsymbol{s}_{\boldsymbol{k}_{\parallel}}$, the momentum-dependent spin polarization is introduced as [40]

$$\boldsymbol{p}_{\boldsymbol{k}_{\parallel}} = \frac{\boldsymbol{s}_{\boldsymbol{k}_{\parallel}}}{\sum_n |\boldsymbol{s}_{n\boldsymbol{k}_{\parallel}}|}, \quad (2)$$

which characterizes both the direction and magnitude of the effective spin polarization associated with propagating states at a given $\mathbf{k}_{\parallel}$.

For a collinear AFM metal, spin is a good quantum number (in the absence of spin-orbit coupling), meaning that the spin state $\boldsymbol{s}_{\boldsymbol{k}_{\parallel}}$ can only have two directions: either along the Néel vector (that is parallel to one of the AFM sublattices) or antiparallel. Its magnitude, $s_{\boldsymbol{k}_{\parallel}} \equiv |\boldsymbol{s}_{\boldsymbol{k}_{\parallel}}| = N^{\uparrow}_{\boldsymbol{k}_{\parallel}} - N^{\downarrow}_{\boldsymbol{k}_{\parallel}}$, represents the difference of the number of conduction channels $N^{\uparrow,\downarrow}_{\boldsymbol{k}_{\parallel}}$ for a given spin up- (↑) and down- (↓) spin electrons at the transverse wave vector $\boldsymbol{k}_{\parallel}$. This implies that the momentum dependent spin polarization is reduced to a scalar $p_{\boldsymbol{k}_{\parallel}} = \frac{N^{\uparrow}_{\boldsymbol{k}_{\parallel}} - N^{\downarrow}_{\boldsymbol{k}_{\parallel}}}{N^{\uparrow}_{\boldsymbol{k}_{\parallel}} + N^{\downarrow}_{\boldsymbol{k}_{\parallel}}}$ [71].

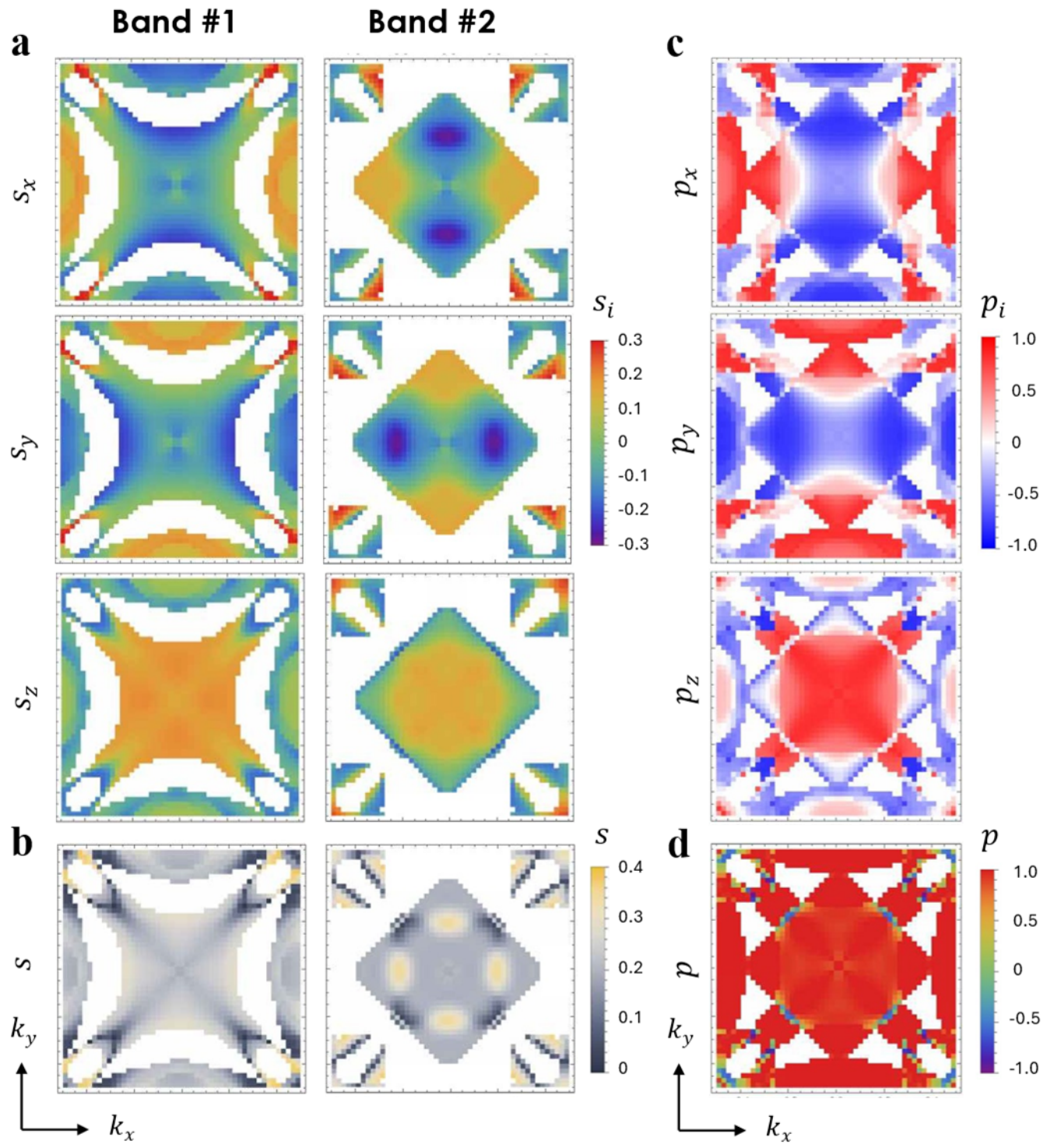

**Fig. 2. Momentum-dependent spin-polarization of bulk $Mn_3NiN$ (001).** (a,b) Two Fermi-surface sheets, labeled bands #1 and #2, projected onto the (001) plane with the corresponding spin-expectation texture, $\boldsymbol{s}_{n\boldsymbol{k}_{\parallel}}$, shown by color for the Cartesian components $s_x$, $s_y$, and $s_z$ (a) for the spin magnitude $s = |\boldsymbol{s}_{n\boldsymbol{k}_{\parallel}}|$ (c,d) Cartesian components $p_x$, $p_y$, and $p_z$ of the effective momentum-dependent spin polarization $\boldsymbol{p}_{\boldsymbol{k}_{\parallel}}$ (c) and its magnitude $p = |\boldsymbol{p}_{\boldsymbol{k}_{\parallel}}|$ (d).

In noncollinear antiferromagnets, however, spin is generally not aligned along a fixed global axis, and hence both the magnitude and direction of $\boldsymbol{p}_{\boldsymbol{k}_{\parallel}}$ become essential. This behavior is illustrated in Fig. 2 for $Mn_3NiN$ (001). The Fermi surface of bulk $Mn_3NiN$ is primarily composed of two bands. For each band, the spin state $\boldsymbol{s}_{\boldsymbol{k}_{\parallel}}$ exhibits a complex $\boldsymbol{k}_{\parallel}$ dependence, with all three spin components and the spin magnitude $s = |\boldsymbol{s}_{n\boldsymbol{k}_{\parallel}}|$ varying across the 2DBZ, as shown in Figs. 2(a,b). As a result, the effective spin polarization vector $\boldsymbol{p}_{\boldsymbol{k}_{\parallel}}$ adopts different orientations, as reflected in its Cartesian components displayed in Fig. 2(c).

Remarkably, we find that the magnitude of spin polarization $|\boldsymbol{p}_{\boldsymbol{k}_{\parallel}}|$ approaches 100% over a large portion of the Brillouin zone for $Mn_3NiN$ (001) [Fig. 2(d)]. Such complete spin polarization is expected in regions where only one conduction channel is available, for which definition (2) yields $|\boldsymbol{p}_{\boldsymbol{k}_{\parallel}}| = 1$. In the multiband case, however, other highly polarized regions in the 2DBZ arise more accidentally from similarities in the spin textures of the bands forming the Fermi surface, rather than being strictly enforced by the magnetic symmetry. The large effective momentum dependent spin polarization is a prerequisite for achieving a sizable TMR in the AFMTJs based on $Mn_3NiN$ (001) electrodes.

### C. $Mn_3NiN/LaAlO_3/Mn_3NiN$ (001) AFMTJ: Atomic structure and electronic properties

To explore spin-dependent tunneling originating from noncollinear $Mn_3NiN$, we design $Mn_3NiN/LaAlO_3/Mn_3NiN$ (001) AFMTJ as shown in Fig. 3(a) (see also Sec. II). The choice of perovskite oxide $LaAlO_3$ as an insulating barrier is motivated by its large bands gap and a close match of crystal symmetries and lattice constants of $Mn_3NiN$ and $LaAlO_3$. This structural compatibility supports epitaxial growth [72,73], enabling high-quality interfaces that are essential for conserving the transverse crystal momentum $\boldsymbol{k}_\parallel$ during tunneling.

The layer-resolved density of states (LDOS) of $Mn_3NiN$ three unit cells away from the $AlO_2$-$Mn_2N$ interface [B-MNN, Fig. 3(b)] closely resembles the bulk DOS [Fig. A1(d)], indicating that bulk-like electronic properties are recovered away from the interface. The corresponding spectral density ($\boldsymbol{k}_\parallel$-resolved LDOS) reveals how the states at the Fermi energy are distributed across the 2DBZ and shows the expected dominant contribution from Mn $d$-orbitals, with only minor $p$-orbital and negligible $s$-orbital character [Fig. A3(a)].

The LDOS of the central region of the $LaAlO_3$ barrier [B-LAO, Fig. 3(d)] exhibits a band gap of about 3.1 eV, closely matching that of bulk $LaAlO_3$ [Fig. A2(a)]. The Fermi energy of the heterostructure lies well within this band gap, about 1.48 eV above the valence band minimum (VBM) of $LaAlO_3$, confirming that transport across the AFMTJ occurs via quantum tunneling [Fig. 3(d)]. The corresponding spectral density at the Fermi energy reflects the $\boldsymbol{k}_\parallel$-resolved distribution of metal-induced gap states within the barrier [Fig. 3(h)]. Notably, regions of enhanced spectral density correlate with transverse wave vectors for which tunneling is strongest. Orbital decomposition of the spectral density further reveals that these enhanced regions are predominantly associated with the O $p$-orbitals localized on the $AlO_2$ layers [Fig. A3(c)].

At the interfaces, additional electronic features emerge. For the $LaAlO_3$ interfacial layers (I-LAO), the LDOS reveals a broad resonant state slightly below the conduction band minimum, producing a peak at around 1 eV [Fig. 3(e)]. This peak originates from hybridization between O $p$-orbitals of the $AlO_2$ layer and Mn $d$-orbitals of the adjacent $Mn_2N$ layer. The presence of this interfacial hybridization is further evidenced by a redistribution of the spectral density of the $AlO_2$ layer away from the center of the 2DBZ, as shown in Fig. 3(i) [compare to Fig. 3(h)].

On the $Mn_3NiN$ side, the interfacial $Mn_2N$ layer exhibits a modest increase of the LDOS compared to bulk-like layers [Fig. 3(c)]. This increase is due to the polar nature of the $LaAlO_3$ (001) surface, which is terminated by a negatively charged $AlO_2^{-1}$ layer. This leads to the charge redistribution at the interface, as the $AlO_2^{-1}$ layer donates electrons to the adjacent $Mn_2N$ layer. This interfacial charge transfer leads to the appearance of an ×-shaped pattern along the diagonals of the square 2DBZ in the spectral density of the interfacial $Mn_2N$ layer [Fig. 3(g)], a feature that is absent in the bulk-like $Mn_2N$ layers [Fig. 3(f)].

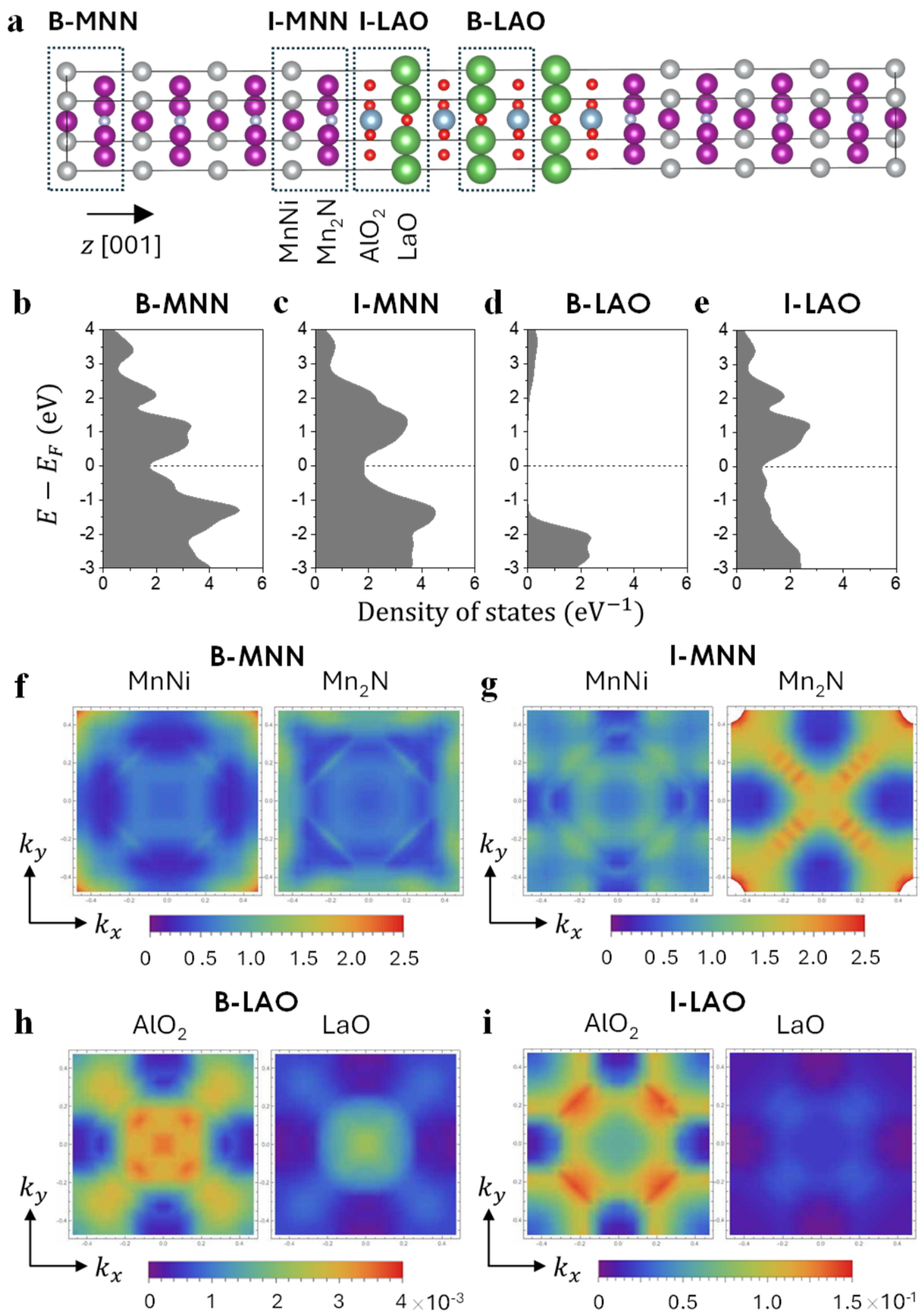

**Fig. 3. Atomic structure and electronic properties of $Mn_3NiN/LaAlO_3/Mn_3NiN$ (001) AFMTJ.** (a) Atomic structure of the AFMTJ with a lowest-energy interface termination $Mn_2N$-$AlO_2$. Dashed boxed indicate four representative regions of the AFMTJ: bulk-like $Mn_3NiN$ (B-MNN), interfacial $Mn_3NiN$ (I-MNN), bulk-like $LaAlO_3$ (B-ALO), and interfacial $LaAlO_3$ (I-ALO). (b-e) Calculated layer-resolved density of states (LDOS) of bulk-like $Mn_3NiN$ (b), interfacial $Mn_3NiN$ (c), bulk-like $LaAlO_3$ (d), and interfacial $LaAlO_3$ (e), corresponding to the regions highlighted in panel (a). The dashed line indicated the Fermi energy. (c) Orbital-resolved spectral density at the Fermi energy for of bulk-like $Mn_3NiN$ (f), interfacial $Mn_3NiN$ (g), bulk-like $LaAlO_3$ (h), and interfacial $LaAlO_3$ (i), corresponding to the regions highlighted in panel (a). Each panel is further decomposed into contributions from the two constituent sublayers: MnNi and $Mn_2N$ in (f,g) and $Al_2O$ and LaO in (h,i). The spectral densities are given in units of $\left(\frac{a}{2\pi}\right)^2 \frac{1}{eV}$.

### D. $Mn_3NiN/LaAlO_3/Mn_3NiN$ (001) AFMTJ: Transport properties

Transmission of the $Mn_3NiN/LaAlO_3/Mn_3NiN$ (001) AFMTJ is calculated as described in Methods (Sec. II). For the parallel

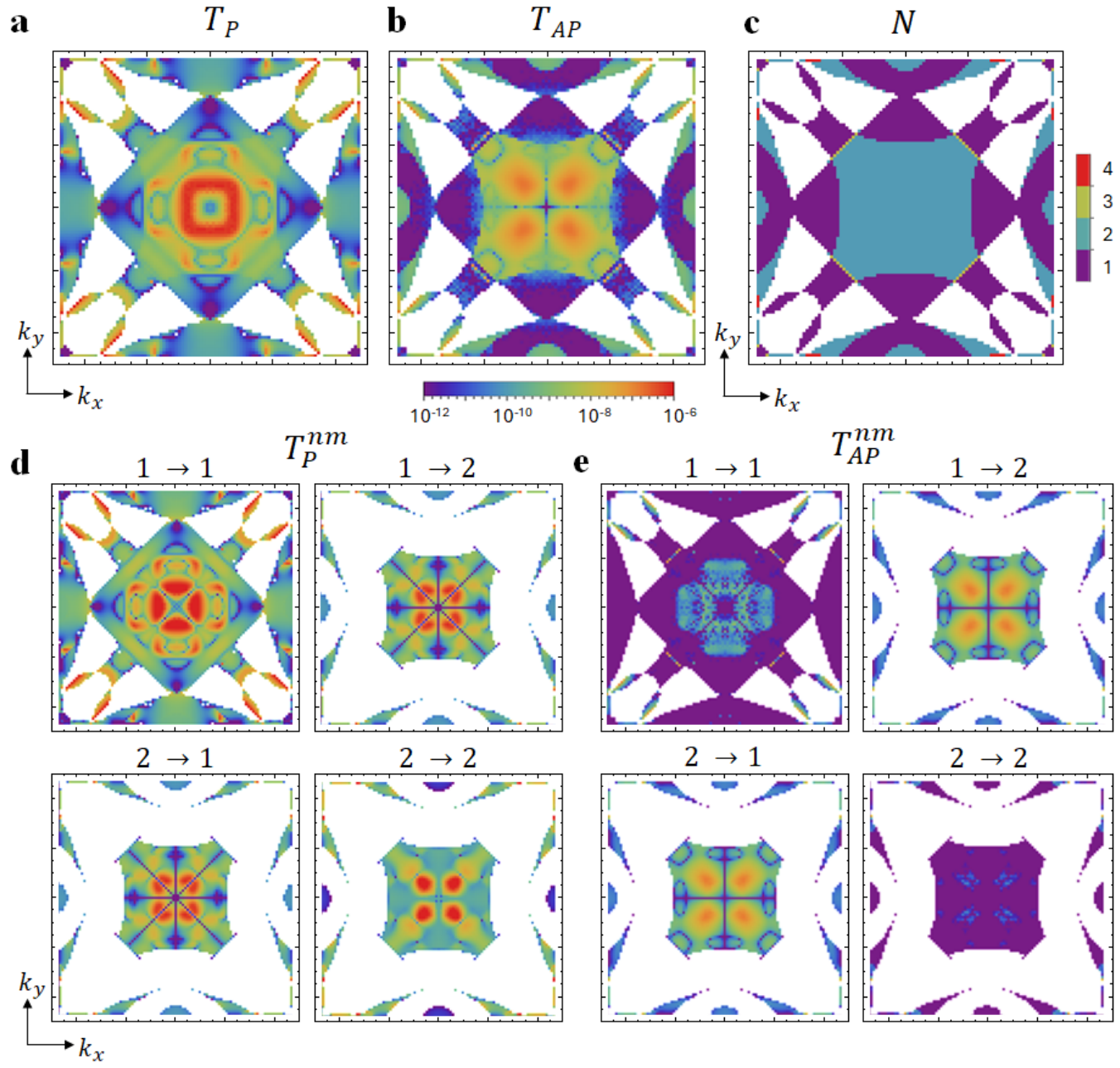

**Fig. 4. Calculated $\boldsymbol{k}_{\parallel}$-resolved transmission.** (a,b) Transmission of the $Mn_3NiN/LaAlO_3/Mn_3NiN$ (001) AFMTJ for parallel ($T_P$) (a) and antiparallel ($T_{AP}$) (b) Néel-vector configurations. (c) Conduction channels $N$ available for transmission in bulk $Mn_3NiN$ (001). (d,e) Decomposition of the transmission into band-to-band contributions for the parallel (d) and antiparallel (e) configurations. The contributing bands are labeled by indices 1 and 2.

configuration, where the Neél vectors of the two magnetic electrodes are parallel, the calculated transmission $T_P$ is $2.31\times10^{-4}$, while for the antiparallel state, where the Neél vectors are antiparallel, the transmission $T_{AP}$ is $1.06\times10^{-5}$ giving rise to a TMR ratio $\frac{T_P - T_{AP}}{T_{AP}}$ of 2075%. This large TMR reflects a large spin polarization of $Mn_3NiN$ (001) [Fig. 1(f)]. Notably, TMR remains robust against shifts of the Fermi energy, with the maximum value approaching 4500%, as shown in Fig. A4(b).

To elucidate these results, we calculated $\boldsymbol{k}_{\parallel}$ -resolved transmission across the 2DBZ of the AFMTJs. Fig. 4(a) shows the results for $T_P(\boldsymbol{k}_{\parallel})$. As expected, the overall transmission distribution pattern in the 2DBZ mirrors the distribution of conduction channels in bulk $Mn_3NiN$ (001) shown in Fig. 4(c). The dominant contribution to the transmission originates from a rounded-square region surrounding the center of the 2DBZ (the $\bar{\Gamma}$ point) while the transmission is somewhat suppressed exactly at the $\bar{\Gamma}$ point (note the logarithmic scale in the plot).

Fig. 4(b) presents the corresponding results for $T_{AP}(\boldsymbol{k}_{\parallel})$. In this configuration, the overall transmission is significantly reduced compared to $T_P(\boldsymbol{k}_{\parallel})$, and the transmission at the $\bar{\Gamma}$ point is completely suppressed. The largest contributions instead arise from four petal-shaped regions surrounding the $\bar{\Gamma}$ point.

Since two dominant conduction bands (channels) (labeled 1 and 2) are present in each electrode, we decompose the transmission into partial band-to-band contributions $T^{nm}(\boldsymbol{k}_{\parallel})$ $(m, n = 1{,}2)$ so that the total transmission is given by $T(\boldsymbol{k}_{\parallel}) = \sum_{nm} T^{nm}(\boldsymbol{k}_{\parallel})$ . The resulting decomposition is shown in Fig. 4(d) for the P configuration and in Fig. 4(e) for the AP configuration.

The first notable feature revealed by this decomposition is a strong suppression of the AP transmission compared to the P transmission for intraband transitions $1(2) \rightarrow 1(2)$ across the entire 2DBZ. This behavior is evident from a comparison of the corresponding intraband transmission profiles in Figs. 4(d) and 4(e). This reduction of the AP transmission is driven by spin mismatch between corresponding conduction channels in the two magnetic electrodes, arising from their opposite spin orientations.

Several additional features in the band-resolved transmission merit discussion. First, the transmission at the $\bar{\Gamma}$ point is significantly reduced in the P configuration of the AFMTJ. Then, as shown in Fig. 4(d) for the P configuration of the AFMTJ, the intraband transmission $T_{\mathrm{P}}^{11}$ exhibits an $\times$-shaped suppression pattern along the diagonals of the square 2DBZ, while $T_{\mathrm{P}}^{22}$ displays an $+$-shaped suppression pattern along the $k_x$ and $k_y$ axes. In both cases, the transmission is significantly reduced in the corresponding regions of the 2DBZ. By contrast, the interband transmissions $T_{\mathrm{P}}^{12}$ and $T_{\mathrm{P}}^{21}$ are vanishing across both the $\times$- and $+$-shaped regions. In the AP configuration, shown in Fig. 4(e), a qualitatively different behavior emerges: the interband transmissions $T_{\mathrm{AP}}^{12}$ and $T_{\mathrm{AP}}^{21}$ are suppressed only within the $+$-shaped region, while remaining sizable along the diagonals of the 2DBZ.

These features in the transmission patterns are governed by band symmetries of the propagating Bloch states in bulk $Mn_3NiN$ and the evanescent states in bulk $LaAlO_3$.

### E. Band symmetries of bulk $Mn_3NiN$ and $LaAlO_3$ and their effects on transmission

To elucidate the features observed in the transmission spectra, we examine the relevant band symmetries of bulk $Mn_3NiN$ and $LaAlO_3$. Since these transmission features are controlled primarily by the orbital character of the propagating and evanescent states, rather than by spin matching alone, we analyze them using the underlying nonmagnetic crystallographic symmetries as a convenient reference framework for classifying the relevant orbital selection rules. For transport along the [001] direction, the transverse wave vector $\boldsymbol{k}_{\parallel}$ lies in the $k_x$-$k_y$ plane, while the electron transmission occurs along the $k_z$ direction [Fig 5(a)]. At the $\bar{\Gamma}$ point ($\boldsymbol{k}_{\parallel} = 0$) of the 2DBZ, transport proceeds along the high-symmetry $\Gamma$–Z line of the three-dimensional Brillouin zone (3DBZ). Along this $\Gamma$–Z line, the two bands forming the Fermi surface of $Mn_3NiN$ [Fig. 5(b), left panel] possess a symmetry-enforced, doubly degenerate $\Delta_5$ symmetry character which belongs to the little group $C_{4v}$. As evident from Figs. A1(b,c), this doubly degenerate band is composed of states with the $d_{xz}$ and $d_{yz}$ orbital character.

Away from $\bar{\Gamma}$, for $\boldsymbol{k}_\parallel = (k_x, 0)$ or $(0, k_y)$, which define the $+$-shaped region in the 2DBZ, the relevant electronic states lie along $(k_x, 0, k_z)$ and $(0, k_y, k_z)$ lines. Along these directions, the little-group symmetry is reduced from that of the Γ–Z line to $C_s$. As a result, the $\Delta_5$ doublet splits into two branches with opposite mirror parity. We denote these branches by $A_1$ and $A_2$ corresponding to states that are even and odd, respectively, under the relevant mirror plane $M_s$. Here $M_s = M_y$ for $(k_x, 0, k_z)$ and $M_s = M_x$ for $(0, k_y, k_z)$ [Fig. 5(b), middle panel].

An analogous symmetry reduction occurs along the diagonal directions of the 2DBZ. For $\boldsymbol{k}_\parallel = (k_x, k_x)$ or $\boldsymbol{k}_\parallel = (k_x, -k_x)$, which define the $\times$-shaped region in the 2DBZ, the relevant electronic states lie along the $(k_x, k_x, k_z)$ and $(k_x, -k_x, k_z)$ lines. Along these directions, the little-group symmetry is likewise $C_s$, but with a different mirror plane. In this case, the $\Delta_5$ representation splits into two components, denoted $B_1$ and $B_2$ which are even and odd, respectively, under the relevant mirror operation $M_s$, where $M_s = M_{(1\bar{1}0)}$ for $k_x = k_y$ and $M_s = M_{(110)}$ for $k_x = -k_y$ [Fig. 5(b), right panel].

These symmetry properties of the propagating Bloch states in $Mn_3NiN(001)$ account for the observed behavior of the interband transmissions $T_P^{12}$ and $T_P^{21}$, which vanish across both the $\times$- and $+$-shaped contours of the 2DBZ in the P-aligned AFMTJ [Fig. 4(d)]. The two bands involved in the interband transmission belong to opposite symmetry representations, one even and the other odd with respect to the relevant mirror symmetry. As a result, the interband transmission channels vanish identically by symmetry.

Next, we analyze the symmetry and decay rates of the evanescent states in bulk $LaAlO_3$ along the same symmetry directions by connecting the complex bands to the corresponding real Bloch states. The results are shown in the left panel of Fig. 5(f) for the Γ–Z line. It is seen that in the vicinity of the Fermi energy (which is matched to that obtained for the $Mn_3NiN/LaAlO_3/Mn_3NiN$ heterostructure) the lowest decay rates are observed for three evanescent states with $\Delta_1$, $\Delta_2$, and $\Delta_5$ symmetries. These evanescent states originate from the continuation of the corresponding Bloch states into the complex-$k$ plane and therefore retain the same orbital character and symmetry. In particular, the band of $\Delta_1$ symmetry is composed of $d_{z^2}$, $p_z$, and $s$ orbitals, the $\Delta_2$ band is dominated by the $d_{x^2-y^2}$ orbital, and the doubly degenerate $\Delta_5$ band originates from the $(p_x, p_y)$ and $(d_{xz}, d_{yz})$ orbitals [Fig. A2(b)].

It is notable that the evanescent state of the $\Delta_5$ symmetry, which matches the symmetry propagating states in $Mn_3NiN$, exhibits only the third-lowest decay rate at the Fermi energy [Fig. 5(f), left panel]. This mismatch explains the reduced transmission at the $\bar{\Gamma}$ point in the P-aligned AFMTJ [Fig. 4(a)]. Away from the $\bar{\Gamma}$ point, this constraint is lifted as the band symmetry is reduced, allowing transmission to proceed through evanescent states with lower decay rates [Fig. A2(c)]. This leads to enhanced transmission in the rounded-square region surrounding the $\bar{\Gamma}$ point [Fig. 4(a)].

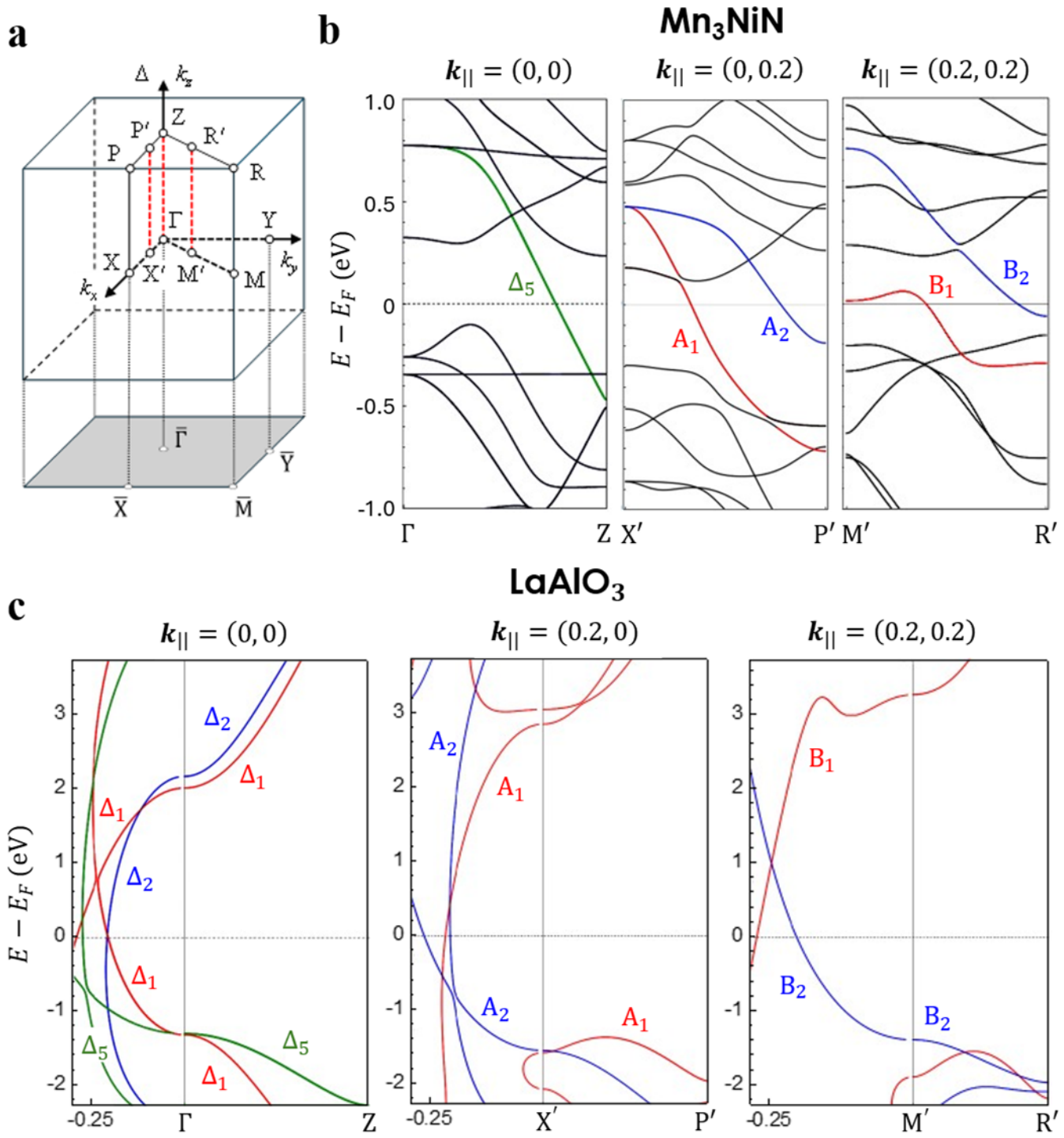

**Fig. 5. Symmetry of representative real and complex bands in bulk $Mn_3NiN$ and $LaAlO_3$.** (a) Schematic illustration of the bulk Brillouin zone and its projection onto the (001) two-dimensional Brillouin zone (2DBZ). High-symmetry points are indicated together with the three lines (highlighted in red) along which the band structures in panels (b) and (c) are calculated. (b) Electronic structure of $Mn_3NiN$ calculated along the lines Γ–Z (left panel), X′ (0.2, 0, 0) – P′ (0.2, 0, 0.5) (central panel), and M′ (0.2, 0.2, 0) – R′ (0.2, 0.2, 0.5) (right panel). Symmetry of the bands crossing the Fermi energy (horizontal dashed line) are indicated. (c) Complex and real band structure of bulk $LaAlO_3$ calculated along the same lines as in panel (b). The symmetries of the bands closest to the Fermi energy (horizontal dashed line) are indicated. In each panel, the vertical dashed line separates the decay rate $\kappa$ (left) from the real bands (right). The Fermi energy is aligned relative to the $LaAlO_3$ band gap to match its position in the $Mn_3NiN/LaAlO_3/Mn_3NiN$ heterostructure.

Away from $\bar{\Gamma}$, a symmetry analysis analogous to that used for $Mn_3NiN$ applies to $LaAlO_3$. For transverse wave vectors defining the $+$- and $\times$-shaped contours of the 2DBZ, the relevant evanescent states are defined along lines parallel to the $k_z$ direction, and the little-group symmetry is reduced from that of the Γ–Z line to $C_s$. Consequently, the corresponding complex bands can be classified according to their parity with respect to the relevant mirror planes. This classification yields even- and odd-parity evanescent states whose decay rates govern the symmetry-selective tunneling through the barrier.

These properties are illustrated in Fig. 5(c) where we display the decay rates of the evanescent states along two representative lines, $P_1: (0.2,0,0) \rightarrow P_1': (0.2,0,0.5)$ (middle panel) and $P_2: (0.2,0.2,0) \rightarrow P_2': (0.2,0.2,0.5)$ (right panel). It is seen that at the Fermi energy, the odd-parity evanescent state exhibits the slowest decay rate in the $LaAlO_3$ barrier in both cases. This identifies the dominant tunneling channel and underpins the symmetry-selective transmission features discussed below.

Specifically, the $+$ -shaped contour observed in the intraband transmission $T_P^{11}$ and the $\times$-shaped contour observed in $T_P^{22}$ reflect the parity properties of the corresponding bands. Band #1 exhibits even parity along the axial directions $k_\parallel = (k_x, 0)$ and $(0, k_y)$, whereas band #2 exhibits even parity along the diagonal directions $k_x = \pm k_y$. Since the odd-parity evanescent state exhibits the slowest decay rate in $LaAlO_3$, transport along the axial directions is suppressed for $T_P^{11}$, while transport along the diagonal directions is suppressed for $T_P^{22}$.

To explain the features observed in the interband transmission patterns of the AP-aligned AFMTJ [Fig. 4(e)], we invoke the mirror symmetry $M_{(1\bar{1}0)}$, which connects the Bloch states of the two oppositely oriented $\Gamma_{4g}$ magnetic configurations. Along the diagonal directions of the 2DBZ, defined by $k_x = \pm k_y$, the transverse momentum remains unchanged under the exchange $k_x \leftrightarrow k_y$. As a result, the mirror symmetry maps each Bloch state onto a symmetry-related state at the same transverse wave vector $k_\parallel$. In the $\Gamma_{4g}$ magnetic structure, the reversal of Néel vector is equivalent to an $M_{(1\bar{1}0)}$ operation [Fig. 1(b)]. Therefore, in the AP-aligned AFMTJ, this symmetry relation allows propagating states in opposite electrodes to acquire compatible orbital character, lifting the interband selection rule present in the P configuration and enabling interband transmission along the diagonal directions of the 2DBZ.

In contrast, along the $k_x = 0$ and $k_y = 0$ directions, the interband transmission remains forbidden even in the AP configuration. Along these lines, the Bloch states are eigenstates of the mirror symmetries $M_y$ and $M_x$, respectively, and the relevant interband states possess opposite mirror parity. Because the tunneling Hamiltonian preserves these mirror symmetries, the corresponding interband matrix elements vanish identically. As a result, the selection rules remain operative in the AP configuration, leading to suppressed transmission along the horizontal and vertical directions of the 2DBZ.

These symmetry selection rules directly account for the transmission pattern observed in the AP-aligned AFMTJ. While the mirror symmetries $M_y$ and $M_x$ continue to suppress interband transmission along the horizontal and vertical directions of the 2DBZ, the diagonal mirror symmetry $M_{(1\bar{1}0)}$ permits interband coupling along $k_x = \pm k_y$. As a result, sizable transmission survives only along the $\times$-shaped contour in the 2DBZ, in contrast to the P-aligned configuration. Consequently, despite the nearly 100% spin polarization of the noncollinear $Mn_3NiN$ (001) electrodes, the AP-aligned junction supports enhanced transmission channels along the diagonals of the 2DBZ, increasing the AP conductance and thereby reducing the resulting TMR.

## IV. DISCUSSION

In the absence of SOC, the transmission patterns are invariant under a rigid rotation of all magnetic moments. In this case, Hamiltonian possesses global spin-rotation symmetry, implying that a uniform rotation of the noncollinear magnetic texture can be removed by a corresponding unitary transformation in spin space. Since the orbital part of the Bloch wave functions and the crystal momentum remain unchanged, the symmetry classification of the propagating states and the associated tunneling selection rules are preserved. Consequently, the $k_\parallel$-resolved transmission maps, including the diagonal interband transmission channels and the suppressed transmission along $k_x = 0$ and $k_y = 0$, remain unchanged under a global rotation of the magnetic moments. For example, the $\Gamma_{5g}$ phase is related to the $\Gamma_{4g}$ phase by a rigid 90° rotation of all magnetic moments within the (111) plane. Therefore, in the absence of SOC the two magnetic configurations are symmetry equivalent and exhibit the same $\boldsymbol{k}_\parallel$-resolved transmission patterns.

It is instructive to compare the results obtained here for AFMTJs based on the noncollinear $\Gamma_{4g}$ antiferromagnet $Mn_3NiN$ with previous studies of AFMTJs based on the noncollinear $\Gamma_{5g}$ antiferromagnet $Mn_3GaN$ [40]. In both cases, the large momentum-dependent spin polarization is responsible for the large TMR. However, in contrast to $Mn_3GaN$, the $Mn_3NiN$-based AFMTJs exhibit enhanced interband transmission in the AP configuration, which increases the AP conductance and thereby reduces the TMR.

This observation indicates that the momentum-dependent spin polarization defined by Eq. (2), while providing useful insight into the possible magnitude of TMR, is not by itself a sufficient descriptor of tunneling transport in noncollinear AFMTJs. In particular, the tunneling conductance is also strongly influenced by symmetry-imposed selection rules and by the orbital character of the propagating Bloch states. In $Mn_3NiN$, the electronic band permits additional interband transmission channels in the AP configuration, whereas these channels remain suppressed in $Mn_3GaN$. Thus, despite similarly large spin polarizations, the detailed magnetic symmetry of the noncollinear state plays a decisive role in determining the actual TMR magnitude.

## V. CONCLUSIONS AND OUTLOOK

In conclusion, we have shown that band symmetry plays an important role in spin-dependent tunneling through crystalline AFMTJs. Using $Mn_3NiN/LaAlO_3/Mn_3NiN$ (001) junctions as a representative example, we demonstrated that although a large momentum-dependent spin polarization of noncollinear AFM electrodes is a prerequisite for achieving sizable TMR, it is not sufficient to fully describe the transport behavior. Instead, the band symmetry of the electrode Bloch states together with their symmetry-selective coupling to evanescent states in the tunneling barrier controls which transmission channels are allowed or suppressed.

We find that symmetry constraints strongly influence interband tunneling in both magnetic configurations. In the parallel configuration, mirror-symmetry selection rules suppress interband transmission, limiting the conductance to intraband channels. In the antiparallel configuration, however, symmetry-related Bloch states can couple across the junction, activating additional interband transmission channels that would otherwise be forbidden. The resulting enhancement of the antiparallel conductance reduces the TMR relative to predictions based solely on momentum-dependent spin polarization. Nevertheless, the calculated TMR remains exceptionally large, exceeding 2000%, underscoring the robustness of the effect in these AFMTJs and demonstrating that magnetic symmetry determines the attainable magnitude of TMR rather than its existence.

Overall, our results establish symmetry-selective tunneling as a fundamental transport mechanism in AFMTJs, placing them on the same conceptual footing as crystalline ferromagnetic MTJs, where symmetry filtering between electrode and barrier states governs the tunneling conductance. At the same time, the present work reveals a key distinction unique to noncollinear antiferromagnets, where the symmetry and orbital matching of electronic bands in the electrodes and tunneling barrier interplay with the noncollinear magnetic structure.

Extending this analysis to other noncollinear antiferromagnets and to barriers with different crystal symmetries and orbital characters, such as $SrTiO_3$, MgO, or ferroelectric oxides, opens new opportunities for engineering TMR through controlled band matching and symmetry filtering. Overall, these results demonstrate that the symmetry and orbital character of Bloch and evanescent states provide a powerful framework for understanding and designing spin-dependent tunneling phenomena in AFM spintronic devices.

## ACKNOWLEDGMENTS

The authors thank Gautam Gurung for his help in transport calculations. This work was supported by the National Science Foundation (NSF) through the Future of Semiconductors (FuSe) program under Award No. 2425567. Computations were performed at the University of Nebraska Holland Computing Center.

## DATA AVAILABILITY

The data that support the findings of this paper are not publicly available. The data are available from the authors upon reasonable request.

### APPENDIX I: Band Structure of Bulk $Mn_3NiN$

Fig. A1 shows the orbital-resolved band structure of $Mn_3NiN$ along the high-symmetry lines Γ-Z-M-Γ-R-Z [Figs. A1(a-c)] together with the density of states around the Fermi energy $E_F$ [Fig. A1(d)]. This band structure reveals that the electronic states near the Fermi energy are predominantly derived from Mn $d$ orbitals. The dispersive bands crossing $E_F$ along the Γ-Z line are composed of the $d_{xz}$ and $d_{yz}$ orbitals and are largely responsible for the transport properties of $Mn_3NiN$.

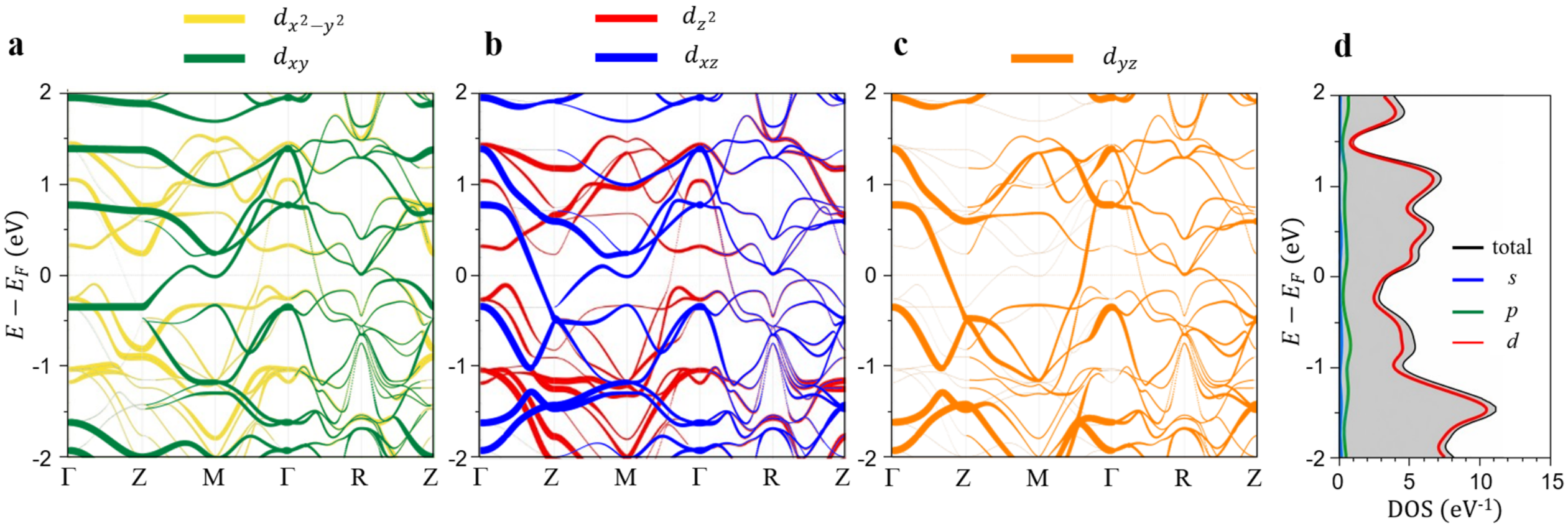

**Fig. A1. Orbital-resolved electronic band structure and density of states (DOS) of bulk $Mn_3NiN$.** (a–c) The band structure along the high-symmetry directions $\Gamma$-Z-M-$\Gamma$-R-Z, with the line thickness proportional to the orbital weight of the corresponding atomic states. (d) The total and orbital-resolved density of states near the Fermi energy $E_F$, with contributions from $s$, $p$, and $d$ orbitals indicated separately.

### APPENDIX II: Band Structure of Bulk $LaAlO_3$

Fig. A2(a) shows the band structure of bulk $LaAlO_3$ calculated using an $f$-in-core approximated pseudopotential. This approach reproduces all essential features of the valence and conduction bands obtained with more sophisticated methods [74,75], except for the well-known underestimation of the band gap within the DFT-GGA approximation. This limitation is not critical for the present study, since our analysis focuses on the valence-band

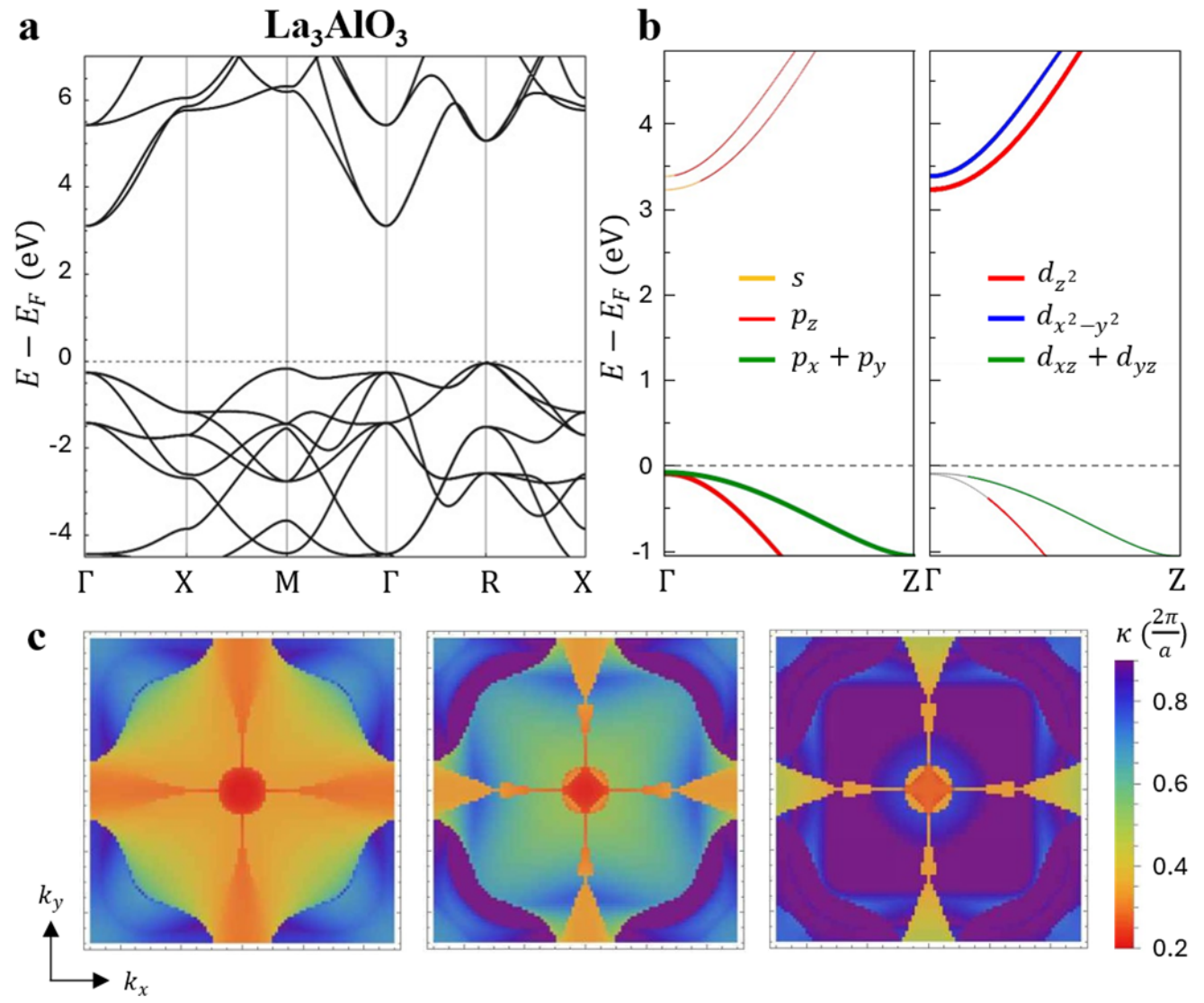

**Fig. A2. Band structure and decay rates of evanescent states in bulk LaAlO₃.** (a) Electronic band structure along the high-symmetry directions of the Brillouin zone. (b) Orbital-resolved band dispersions along the $\Gamma - Z$ line with the marker size proportional to the corresponding orbital weight. (c) $\boldsymbol{k}_{\parallel}$-resolved distributions of the fist (left), second (middle), and third (right) lowest decay rates of the evanescent states in the 2DBZ.

states and does not involve electronic states near or above the conduction-band minimum (CBM).

Fig. A2(b) shows the orbital character of the electronic bands along the Γ-Z line forming the CBM and valence-band maximum (VBM). The orbital character and symmetry of these bands determine the orbital character and symmetry of the corresponding evanescent states shown in Fig. 5(c). The CBM is formed by two bands with Mn $d_{z^2}$ and $d_{x^2-y^2}$ character, related to the $\Delta_1$ and $\Delta_2$ symmetries, respectively. In contrast, the VBM is formed by two bands with dominant $p_z$ and $(p_x, p_y)$ character, corresponding to states of $\Delta_1$ and $\Delta_5$ symmetry. The complex bands shown in Fig. 5(c) originate from the continuation of the corresponding propagating bands into the complex-$k$ plane and therefore retain the same orbital character and symmetry.

Fig. A2(c) shows the momentum-resolved distributions of the first, second, and third lowest decay rates of the evanescent states in the 2DBZ. The slowest-decaying evanescent state exhibits a pronounced minimum of the decay rate around the $\bar{\Gamma}$ point (the center of the 2DBZ), indicating that, in the absence of symmetry constraints, tunneling transport through the barrier is expected to be dominated by states with small transverse wave vector $\boldsymbol{k}_{\parallel}$. In addition to this feature, cross-shaped regions extending along the $k_x$ and $k_y$ directions are clearly visible, reflecting enhanced penetration of evanescent states associated with the $d_{xz}$- and $d_{yz}$-derived bands. The second and third decay channels exhibit progressively larger decay rates and more localized momentum-space features, indicating their reduced contribution to tunneling transport.

## APPENDIX III: LaAlO$_3$/Mn$_3$NiN (001) Interface

Table A2 summarizes the LaAlO$_3$/Mn$_3$NiN interface formation energies, calculated *as* $E_{int} = E_{hs} - E_{sep}$, where $E_{hs}$ is the total energy of the relaxed heterostructure and $E_{sep}$ is the energy of the two constituent slabs separated by a vacuum region of 25 Å to eliminate interlayer interactions. The lowest formation energy, –0.362 meV/u.c., is obtained for the LaO-MnNi interface. However, this structure exhibits an unphysically large residual internal pressure of –2.21 kbar that does not vanish upon structural relaxation. Therefore, the AlO$_2$-Mn$_2$N interface, which has the second-lowest formation energy of –0.301 meV/u.c., was selected for the main analysis presented in this paper. For comparison, transmission results for the AFMTJ with the LaO-MnNi interface are provided in Appendix V.

**Table A1**: Interface formation energy $E_{int}$ and internal pressure $P_{int}$ of different symmetric interfaces between Mn$_3$NiN and LaAlO$_3$.

| Interface | AlO$_2$-MnNi | AlO$_2$-MnNi | LaO-MnNi | LaO-MnNi | AlO$_2$-Mn$_2$N | AlO$_2$-Mn$_2$N | LaO-Mn$_2$N | LaO-Mn$_2$N |
|---|---|---|---|---|---|---|---|---|
| Interface atomic structure | | | | | | | | |
| $P_{int}$ (kbar) | $\lvert P_{int}\rvert < 0.5$ | $\lvert P_{int}\rvert < 0.5$ | $\lvert P_{int}\rvert < 0.5$ | –2.21 | $\lvert P_{int}\rvert < 0.5$ | –0.36 | $\lvert P_{int}\rvert < 0.5$ | $\lvert P_{int}\rvert < 0.5$ |
| $E_{int}$ (meV/u.c.) | –0.288 | –0.174 | –0.178 | –0.362 | –0.238 | –0.301 | –0.164 | –0.145 |

## APPENDIX IV: Orbital-Resolved Spectral Density

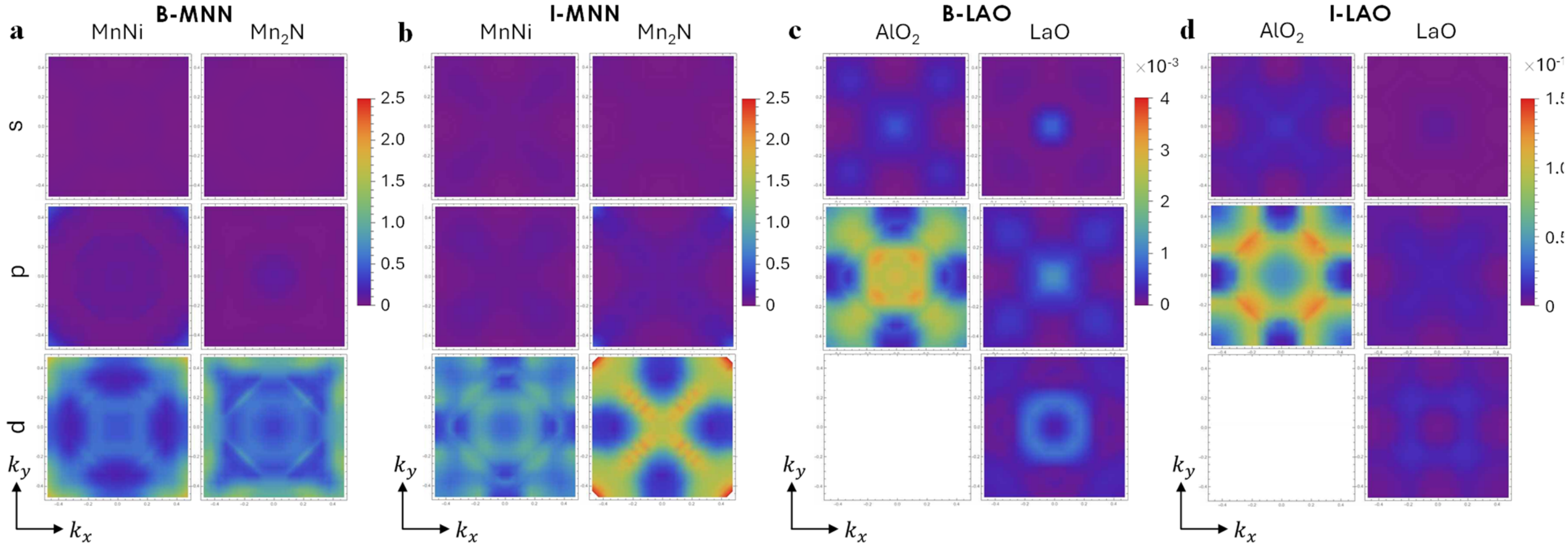

**Fig. A3.** Orbital-resolved spectral density at the Fermi energy for of (a) bulk-like $Mn_3NiN$, (b) interfacial $Mn_3NiN$, (h) bulk-like $LaAlO_3$, and (i) interfacial $LaAlO_3$, corresponding to the regions highlighted in Fig. 3 (a). Each panel is further decomposed into contributions from the two constituent sublayers: MnNi and $Mn_2N$ in (a,b) and $Al_2O$ and LaO in (c,d). The spectral densities are given in units of $\left(\frac{a}{2\pi}\right)^2 \frac{1}{eV}$.

## APPENDIX V: Energy Dependent Transmission and Transmission for the LaO-MnNi Interface

Fig. A4(a) shows TMR as a function of energy for $AlO_2$-$Mn_2N$ terminated $Mn_3NiN/LaAlO_3/Mn_3NiN$ (001) AFMTJ which is the primary focus of this work. Notably, the corresponding TMR ratio [Fig. A4(b)] remains consistently large over the entire energy range considered, varying from about 400% to 4500%.

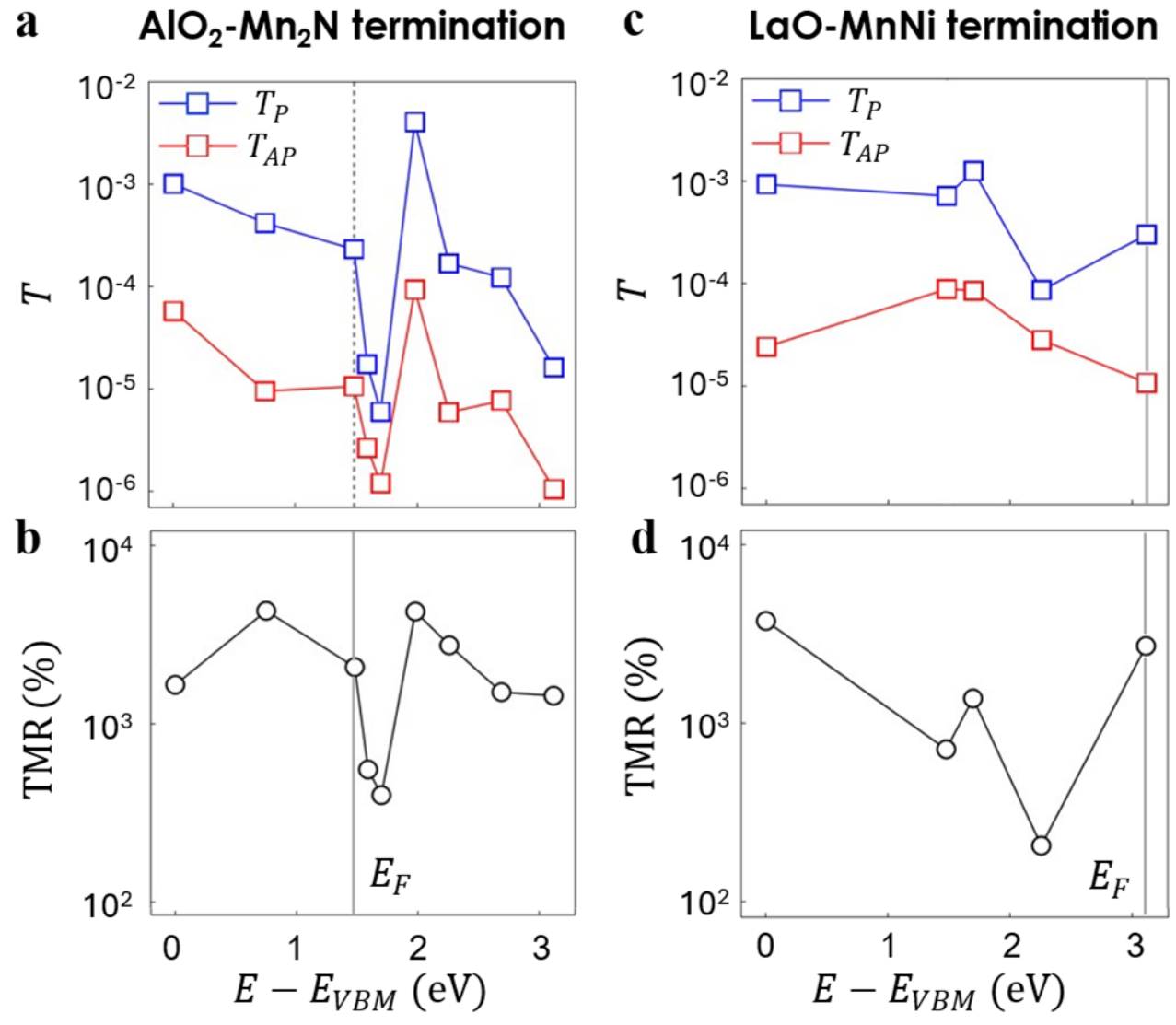

**Fig. A4.** Energy-dependent transmission for the parallel ($T_P$) and antiparallel ($T_{AP}$) Néel-vector configurations, together with the corresponding TMR ratio for $Mn_3NiN/LaAlO_3/Mn_3NiN$ (001) AFMTJs with (a,b) $AlO_2$-$Mn_2N$ and (c,d) LaO-MnNi interface terminations. The transmission and TMR are plotted as a function of energy $E$ relative to the valence-band maximum $E_{VBM}$. The vertical dashed lines indicate the position of the Fermi energy $E_F$ in the corresponding AFMTJs.

For completeness, we have also calculated transmission and TMR for $Mn_3NiN/LaAlO_3/Mn_3NiN$ (001) junctions with LaO-MnNi terminated interfaces. Fig. A5 shows $\boldsymbol{k}_{\parallel}$-resolved transmission for these junctions for parallel ($T_P$) and antiparallel ($T_{AP}$) Néel-vector configurations. Although the results differ quantitatively from those obtained for the $AlO_2$-$Mn_2N$ terminated AFMTJs [Figs. 4(a,b)], the transmission patterns exhibit qualitatively similar behavior. In particular, the parallel configuration shows substantially enhanced transmission compared to the antiparallel configuration over most regions of the 2DBZ. The total transmission of this junction is $3.01\times10^{-4}$ for the parallel configuration and $1.07\times10^{-5}$ for the antiparallel, corresponding to the TMR ratio of about 2700%.

Figs. A4 (c, d) show the transmission and TMR as a function of energy $E$ (with respect the VBM energy $E_{VBM}$) for the LaO-MnNi terminated $Mn_3NiN/LaAlO_3/Mn_3NiN$ (001) AFMTJ. The predicted TMR remains substantial over the considered energy range, varying between approximately 200% and 3800%. These results demonstrate that the predicted TMR is largely insensitive to the interface termination and is governed primarily by the bulk electronic structure of the $Mn_3NiN$ electrodes, as well as by the symmetry matching between the propagating Bloch states in the electrodes and the evanescent states in the $LaAlO_3$ barrier.

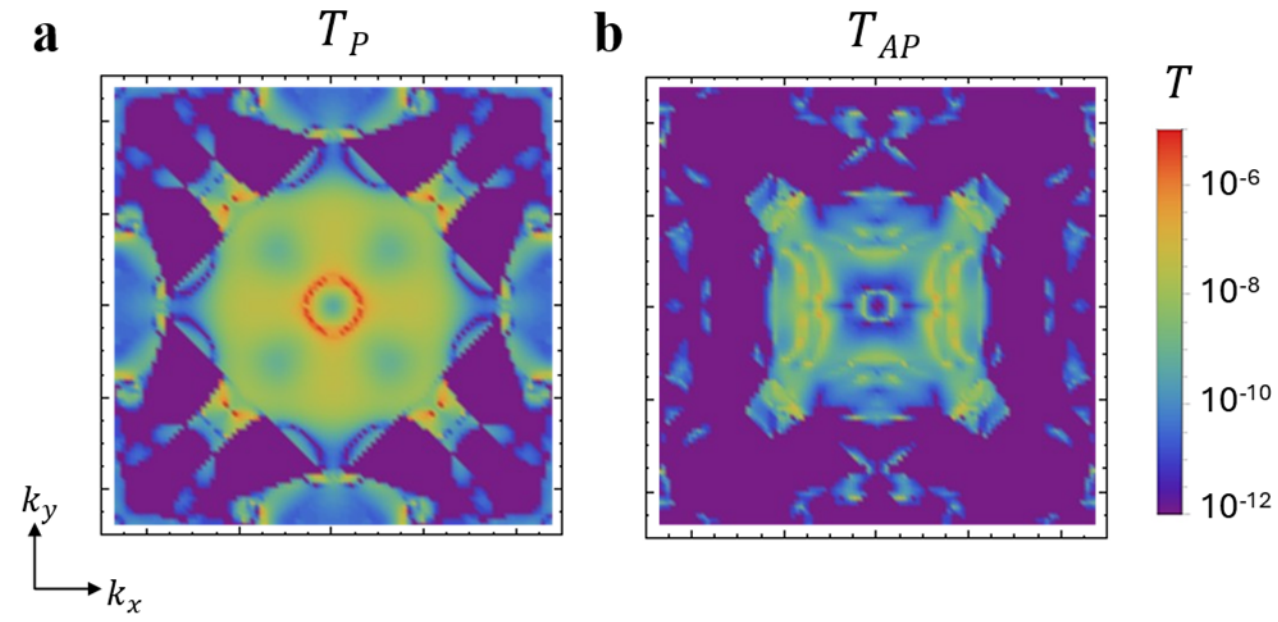

**Fig. A5.** $\boldsymbol{k}_{\parallel}$-resolved transmission of the LaO-MnNi interface-terminated $Mn_3NiN/LaAlO_3/Mn_3NiN$ (001) AFMTJ for parallel ($T_P$) (a) and antiparallel ($T_{AP}$) (b) Néel-vector configurations.